# Influence of some cardiovascular risk factors on the relationship between age and blood pressure

Giulia Silveri[1], Lorenzo Pascazio[2], Miloš Ajčević[1], Aleksandar Miladinović[1], Agostino Accardo[1]

[1]Department of Engineering and Architecture, University of Trieste, Trieste, Italy
[2]Department of Medicine, Surgery and Health Science, University of Trieste, Trieste, Italy
giulia.silveri@phd.units.it

**Abstract.** Blood Pressure (BP) is a biological signal related to the cardiovascular system that inevitably is affected by ageing. Moreover, it is also influenced by the presence of cardiovascular risk factors. To evaluate how the relationship between BP and age changes with the presence of risk factors in hypertensive and normotensive subjects, we analyzed 880 subjects with and without smoking, obesity, diabetes mellitus and dyslipidemia. A regression line fitted each BP/Age relation calculated separately for normotensive and hypertensive subjects with and without risk factors. For each of the four conditions the office and the 24-hour ambulatory BP monitoring (ABPM) were considered. In subjects with and without risk factors, the slopes of the Systolic BP/Age relation were higher in hypertensive than in normotensive subjects in both office and ABPM conditions. Moreover, the presence of risk factors modified the Systolic BP/Age relation in hypertensive subjects by using either office or ABPM measurements. Finally, we confirmed that the difference between the two modalities depends on age too.

## 1      Introduction

One of the strongest high risk factors for mobility and mortality worldwide is the elevated Blood Pressure (BP) [1]. An accurate diagnosis of hypertension is crucial to target treatment to those individuals at high risk for adverse events. For many years, the gold standard for defining the BP status has been the office technique made in the physician's office, although this technique has many potential limitations. The inability of correlating the level of BP both to damage of the target organ and to the risk of cardiovascular disease is minimized using 24-hour ambulatory BP monitoring (ABPM) that can provide a more accurate estimate of BP [2]. Since ABPM values are usually considered lower than corresponding office BP, the diagnostic thresholds for hypertension based on 24-hour ABPM are lower than those for office BP [3]. The difference between these two techniques has been used to identify the so-called withe-coat effect and masked hypertension [4, 5]. In the first case, the BP is elevated in office but it is normal when measured by ABPM. Conversely, masked hypertension refers to untreated patients in whom the BP is normal in the office but is elevated when measured by ABPM.

On the other hand, ageing is considered as one of the natural causes influencing the increase in pressure. An epidemiological study on healthy subjects estimated that Systolic BP (SBP) increases with an average of 140 mmHg by the seventh decade, and that the Diastolic BP (DBP) tends to increase with age but the average value tends to remain flat or to decline after the fifth decade [6]. Franklin et al. in The Framingham Heart Study [7] characterized the age related changes in BP in a sample of 20136 normotensive and untreated hypertensive subjects (50 to 79 years old) highlighting that DBP falls after 60 years old while SBP continual rises. Moreover, the differences between office and ambulatory BP increase progressively both with age and office BP values; office measurements were also higher than ABPM values both in normotensive and hypertensive subjects [7-9]. On the contrary, in young healthy subjects, aged 4-18 years old, ABPM values were most often higher than office BP values and this difference was reduced with ageing [10]. In a meta-analysis on 13 healthy population base cohorts, Ishikawa et al. [11] found that BP office increased with age more steeply than ABPM only after 50 years old for SBP and after 45 years old for DBP, but office BP was lower than ABPM in the youngest. Furthermore, Conen et al. [12] compared individual differences between ABPM and office BP according to 10-year age categories in subjects not taking antihypertensive treatment, founding that office BPs increase from 117 to 149 mmHg (SBP) and from 64 to 82 mmHg (DBP) from the youngest to oldest age. Additionally, they highlighted that the relationship between age categories and diastolic ABPM increased until 50 years old and then it decreased. They also observed that among subjects younger than 50 years old, ABPM were higher than office; for age between 50 to 60 years the ABPM and office were similar, and older than 60 years this relationship was inversed.

Moreover, it is known that BP is correlated with some risk factors of cardiovascular disease such as smoking, obesity, dyslipidemia and diabetic mellitus that generally increase its value [6,13-16]. Although BP values increase due to both ageing and some risk factors, until now the studies have examined the relationship between BP and age grouping together hypertensive and normotensive subjects without quantifying the influence of risk factors on this relation [6-8,10-12]. Since from a clinical point of view it can be useful to understand how the relationship between BP and age is affected by risk factors in people with and without hypertension, in this study we accurately examine how the relation between age and BP changes in subjects presenting or not at least one risk factor. The considered factors were smoking, obesity, dyslipidemia and diabetic mellitus, in subjects whose BP were evaluated in office and ABPM ways.

## 2   Methods

The study population included a total of 880 subjects (356 males, 524 females, aged 65±16 years old) with known or suspected hypertension. This retrospective study was carried out at the Ambulatory of Cardiovascular Pathophysiology (Geriatric Department of the University of Trieste, Trieste, Italy) between May 2016 and July 2018. The inclusion criteria, to enter in this study, were : 1) no clinical or laboratory evidence of secondary arterial hypertension, 2) absence of clinical evidence of hypertension-related complications, 3) an office SBP value between 70mmHg and 260mmHg and an office DBP value between 40mmHg and 150mmHg. The study was performed according to the Declaration of Helsinki and all the subjects signed their informed consent.

The clinical classification of the hypertensive subjects was done according to current international Guidelines using either the office and Ambulatory blood pressure levels [3, 17]. Hence, using the office blood pressure levels, the subjects were identified either as hypertensive if the office SBP≥140mmHg or DBP≥90mmHg or as normotensive if the office SBP<140mmHg and DBP<90mmHg. Then using the Ambulatory BP levels, the subjects were classified as hypertensive if the 24h meanSBP≥130mmHg or 24h meanDBP≥80mmHg or as normotensive if the 24h meanSBP<130mmHg and 24h meanDBP<80mmHg. The information about smoking, obesity, diabetes mellitus and dyslipidemia risk factors were collected during physical examination, in accordance with international guidelines [18]. The subjects were then grouped in eight classes (Table 1) considering them either all together or excluding those presenting at least one risk factor.

**Table 1.** Subjects group

|    | ALL TOGHETER | | WITHOUT RISK FACTORS | |
|----|--------------|------|--------------|------|
|    | OFFICE | ABPM | OFFICE | ABPM |
| H  | 253 | 241 | 54  | 60 |
| NH | 112 | 124 | 105 | 99 |

In order to examine the influence of the presence of risk factors on the relationship between BP and age, the SBP/Age and the DBP/Age relations were separately evaluated for hypertensive and normotensive subjects, by calculating the slope, the intercept and the R-square statistic of a regression line the parameters of the regression lines in each group of subjects.

## 3    Results

Figures 1 and 2 show the behavior of BP/Age relations in hypertensive and normotensive subjects considering office and ambulatory measurements. The parameters (slope and intercept) of the regression line fitting each relationship are reported in Table 2. In particular, in Fig. 1, in which subjects with and without risk factors were considered all together, the intercepts, as expected, were higher in hypertensive than in normotensive subjects and they were 2-8 mmHg greater in office than in ABPM measurements (Table 2).

The SBP/Age slopes were positive in both subject groups with higher values in office condition than in ABPM, while the DBP/Age slopes were negative in all subjects and slightly greater in ABPM than in office (Table 2).

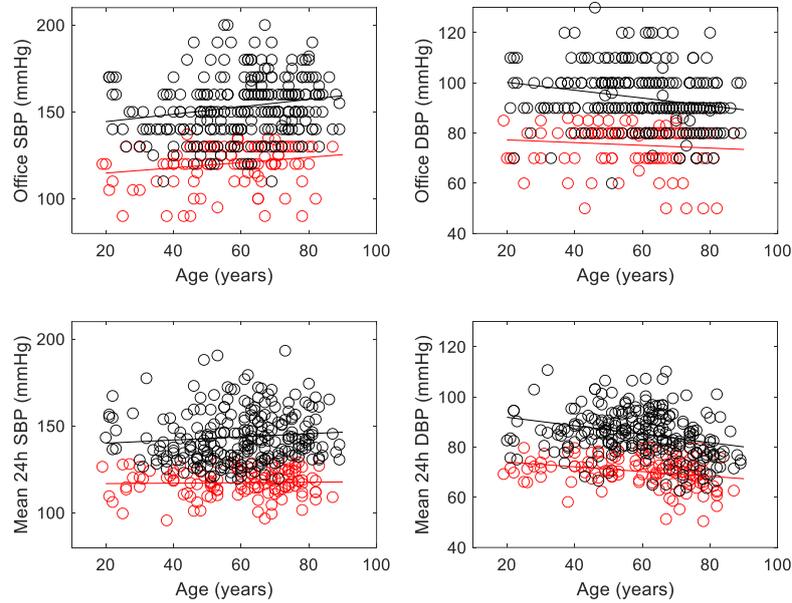

**Fig. 1.** SBP/Age and DBP/Age relationships in all the subjects (black: hypertensive patients; red: normotensive subjects) calculated in office condition (top panels) and as average on 24h (bottom panels).

In all the subjects without risk factors (Fig. 2), the SBP/Age intercepts were comparable in office and ABPM conditions while the DBP/Age intercepts were greater in hypertensive subjects in both conditions. The SBP/Age slopes, in many cases significantly ($p<0.05$) different from zero, were positive in all measurements with the lowest values in normotensive subjects evaluated in ABPM, while the DBP/Age slopes, significantly different from zero in many situations too, presented lower values in office than in ABPM and in normotensive than in hypertensive subjects (Table 2).

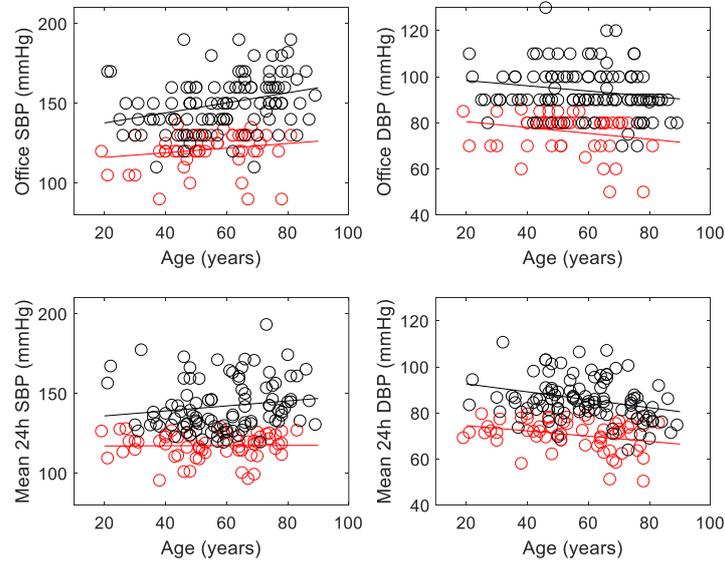

**Fig. 2.** SBP/Age and DBP/Age relationships in the subject without risk factors (black: hypertensive patients; red: normotensive subjects) calculated in office condition (top panels) and as average on 24h (bottom panels).

**Table 2.** Intercepts (mmHg) and slopes (mmHg/year) of the SBP/age and DBP/age relations in office and ABPM measurements either in all the subjects or in subjects without risk factors. N=normotensive subjects; H=hypertensive subjects. * p<0.03.

| **ALL THE SUBJECTS** | | | | | |
|---|---|---|---|---|---|
| | | **OFFICE** | | **ABPM** | |
| | | **N** | **H** | **N** | **H** |
| **SBP/Age** | Intercept | 112 | 140 | 116 | 138. |
| | Slope | 0.15* | 0.22* | 0.01 | 0.09 |
| **DBP/Age** | Intercept | 78 | 103 | 76 | 95 |
| | Slope | -0.05 | -0.16* | -0.10* | -0.17* |
| **SUBJECTS WITHOUT RISK FACTORS** | | | | | |
| | | **OFFICE** | | **ABPM** | |
| | | **N** | **H** | **N** | **H** |
| **SBP/Age** | Intercept | 113 | 131 | 117 | 133 |
| | Slope | 0.15 | 0.32* | 0.005 | 0.157 |
| **DBP/Age** | Intercept | 83 | 101 | 77 | 96 |
| | Slope | -0.13 | -0.11 | -0.12* | -0.17* |

The differences between the regression lines of the SBP/Age relation calculated on all the subjects and on subjects without risk showed, in normotensive subjects (red lines in Fig. 3, left panels), negligible and independent behavior of age while, in hypertensive subjects (black lines) the differences were related to age, decreasing toward zero in elderly.

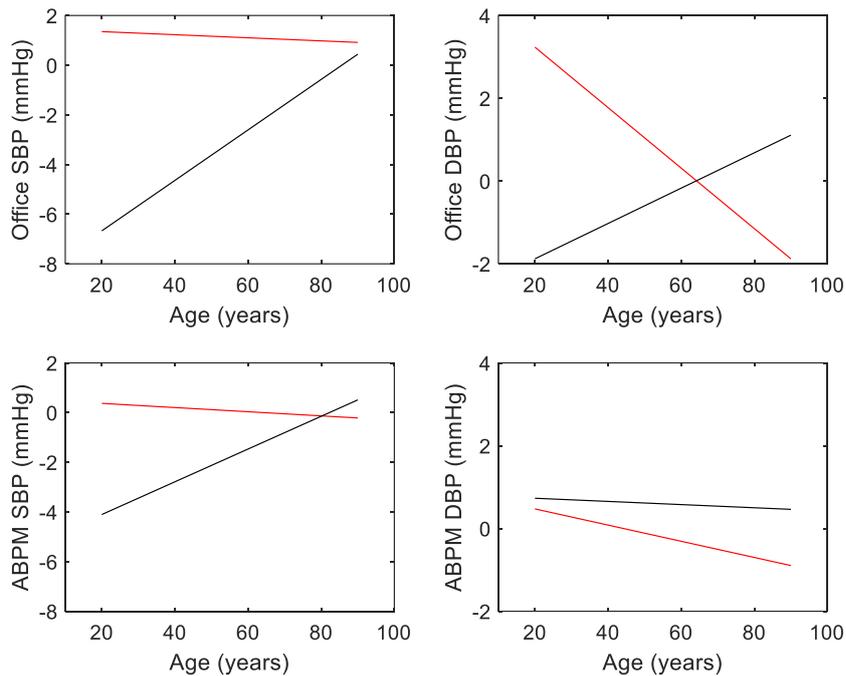

**Fig. 3.** Differences between the regression lines of SBP/Age and DBP/Age relationships calculated on all the subjects and on subjects without risk factors, in office (top panels) and ABPM (bottom panels) modalities. Black lines: hypertensive patients; red lines: normotensive subjects.

These trends were similar for BP calculated both in office and in ABPM ways. On the contrary, the differences between the regression lines of the DBP/Age relation decreased with ageing more remarkable if office instead of ABPM was used, especially in normotensive subjects (red lines in Fig. 3, right panels). In hypertensive subjects (black lines) the differences had trends that were opposite in office and ABPM, presenting a negligible relation with ageing.

The differences between the regression lines of the SBP/Age relation calculated in office and ABPM ways in hypertensive (black lines in Fig. 4, left panels) as well as in normotensive (red lines) subjects increased with age either in all the subjects or in those without risk factors. The differences between the regression lines of the DBP/Age relation increased with ageing too when the subjects were considered all together while, if we consider only those without risk factors, the increase with age is present only in hypertensive subjects (black lines in Fig. 4, right panels).

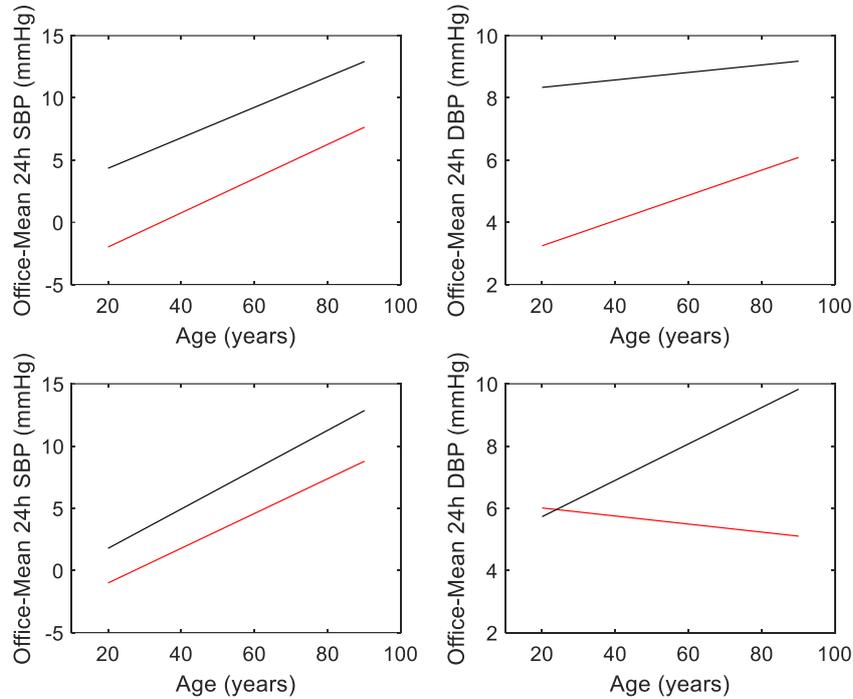

**Fig. 4.** Differences between the regression lines of SBP/Age and DBP/Age relationships calculated in office and ABPM modalities on all the subjects (top panels) and on subjects without risk factors (bottom panels). Black lines: hypertensive patients; red lines: normotensive subjects.

## 4   Discussion

In the literature, several studies evaluated the relation between age and BP measurements but none of them distinguished between subjects with and without the presence of at least one risk factor although it is known that many factors could increase BP values [6,13-15]. Our results about the BP/Age relation, considering either office or ABPM measurements in subjects with and without risk factors, generally confirm the finding of previous researches [7,6,12] although in these studies the subjects were not classified in hypertensive and normotensive. In particular, SBP increased significantly with age (Table 2), presenting different values (between 1.5 and 2.2mmHg per decade) in normotensive and in hypertensive subjects only for office measurements, while for ABPM the increment was negligible. This difference between the two measurement modalities could be partially due to the low accuracy of the office BP frequently rounded to the nearest tens of mmHg. On the other hand, DBP, after a quite constant behavior in the first decades of age, significantly decreases over 50 years, especially for BP measured in ABPM way. According to [19] the increase in SBP with ageing maybe due to a large artery stiffness and to the peripheral vascular resistance in small vessel, instead the increase of DBP up to the age of 50 is due to peripheral vascular resistance and the successive decrease is due to the increase of large artery stiffness.

About the subjects without risk factors, our results highlighted that the dependence on age of SBP in hypertensive subjects, both considering office and ABPM measures, was greater than that assessed considering all subjects together, with decreasing differences with age (black lines in Fig.3, left panels). On the other hand, no differences in normotensive subjects were found (red lines). The DBP/Age relationships were substantially the same considering all the subjects or only those without risk factors when ABPM was used, while differences of opposite sign in the hypertensive and normotensive were present for the office measures (Fig. 3, right panels).

Comparing the differences between the SBP/Age relations obtained measuring the BP in office and in ABPM (Fig.4), a large increase of the differences with age is present in all the situations, partially confirming previous findings [12]. This fact highlights the risk involved in estimating a correct BP when using the office measurement and underlines how ageing increases the error introduced by the measurement method in subject with or without risk factors. The differences in the DBP/Age relations increase with age too, although less than in the SBP/Age except in normotensive subjects without risk factors for which the difference is approximately constant and equal to 5-6mmHg in all the ages considered.

In summary, these results showed that the trend between blood pressure and age in subjects with risk factors was similar to that in subjects without risks, both normotensive and hypertensive. Moreover, we pointed out the presence of differences between subjects with and without risk factors in SBP/Age relation only in hypertensive subjects. The results were similar by using either office or ABPM measurements of SBP confirming, as previously reported in the literature [7-12], the presence of differences between the two modalities quantifying how these changed with age. With this study we would support the use of the ABPM measure rather than office one for a more accurate assessment of BP and knowledge of BP's dependence on age, according to the presence or absence of risk factors, in order to correctly set the diagnosis of hypertension.